\documentstyle[epsfig]{aipproc}

\newcommand{\kint}{\sum_\alpha \,}

\begin{document}
\title{Thermalisation of inhomogeneous quantum scalar fields in
1+1D\thanks{Presented at CAPP 2000 by M. Sall\'e}
}
\author{Mischa Sall\'e \and Jan Smit \and Jeroen C. Vink}
\address{Institute for Theoretical Physics, University of Amsterdam, 
Valckenierstraat 65, 1018 XE Amsterdam, the Netherlands}

\maketitle

\begin{abstract}
Using an improved version of the Hartree approximation, allowing for ensembles
of inhomogeneous configurations, we show in a $\lambda \phi^4$ theory, that
initially the system thermalises with a Bose-Einstein distribution. For later
times and larger couplings we see deviations.
\end{abstract}

\section*{Introduction}
In many parts of high energy physics like early universe physics and heavy ion
collisions one wants to follow quantum field systems in time. Typical
phenomena include reheating after inflation: after a phase of parametric
resonance most energy is in a small part of the spectrum, a non-thermal initial
condition. 

In order to study these systems and to describe phase transitions and thermal
fluctuations between classical minima, non-perturbative approximation schemes
are needed. Examples of high temperature approximations, that have been used
successfully are the classical approximation\cite{grigo,aarts} and the Hartree or
gaussian approximation\cite{mihai}. The latter assumes a gaussian density
matrix, such that all information is contained in the 1- and 2-point functions,
higher point functions can be factorised using Wick's theorem.

\section*{Inhomogeneous Hartree ensemble}
Instead of the commonly used Hartree approximation, which has a spatially
constant mean field we used an improved version, allowing for
\emph{inhomogeneous} configurations. We studied this approximation in a scalar
$\phi^4$ theory in $1+1$ dimensions:
\begin{equation}
{\mathcal L} = - \frac{1}{2} \partial_\mu \hat{\phi} \partial^\mu \hat{\phi} -
\frac{1}{2} \mu^2 \hat{\phi}^2 - \frac{1}{4} \lambda \hat{\phi}^4.
\end{equation}
In the gaussian approximation we can expand the operator field in mode
functions: \nopagebreak[4]
\begin{equation}
\hat{\phi}(x,t) = \phi(x,t) + \kint \left( \hat{a}_\alpha f_\alpha(x,t) +
\hat{a}_\alpha^\dagger f_\alpha^*(x,t) \right),
\end{equation}
where the \emph{time-independent} $\hat{a}_\alpha$ and $\hat{a}_\alpha^\dagger$
satisfy the usual commutation relations. We can choose them such that:
\begin{equation}
\langle \hat{a}_\alpha^\dagger \hat{a}_\alpha \rangle \equiv n_\alpha^0.
\end{equation}
For a free thermal system the $n^0_\alpha$ have a Bose-Einstein form and
the $f_\alpha(x,t)$ are plane waves, with suitable normalisation.

The Heisenberg equations of motions in the gaussian approximation\footnote{The
classical approximation can be obtained by the substitution $3 \rightarrow 0$ in
Eq.~(\ref{eq:eom_phi}) while omitting Eq.~(\ref{eq:eom_mode}). The large~$n$
approximation leads to $3 \rightarrow 1$ in Eq.~(\ref{eq:eom_phi}) and
Eq.~(\ref{eq:eom_mode}). We chose not to use large~$n$ to avoid problems with
would-be Goldstone bosons in $1+1$ dimensions.} become:
\begin{eqnarray}
\mbox{[} \partial_t^2 - \triangle + \mu^2 + \phantom{3} \lambda \phi^2(x,t) +
3 \lambda \Sigma(x,t) \mbox{]} & \phi(x,t) &= 0, \label{eq:eom_phi} \\
\mbox{[} \partial_t^2 - \triangle + \mu^2 + 3 \lambda \phi^2(x,t) +
3 \lambda \Sigma(x,t) \mbox{]} & f_\alpha(x,t) &= 0, \label{eq:eom_mode}
\end{eqnarray}
\begin{equation}
\Sigma(x,t) = \kint |f_\alpha(x,t)|^2 \, (2 n_\alpha^0 + 1).
\end{equation}
We mention that there exists an effective Hamiltonian which leads to the above
equations, suggesting that in the end the system will equilibrate classically
according to this $H_{\rm eff}$. But we hope that this has a timescale much
larger than that of all interesting processes.

\paragraph*{Initial conditions}
We need to specify the initial fields and their conjugate momenta, which
together with the $n_\alpha^0$ specifies an initial gaussian density matrix. By
averaging over different runs with different initial conditions we can build
non-gaussian density matrices, allowing for much more general initial conditions.
Only the time evolution is evaluated with gaussian density matrices. 

We will take the mean field in its zero-temperature minimum, with all energy
in a few of its momentum modes, like after parametric resonance:
\begin{equation}
\phi(x,0) = v \qquad \textrm{and} \qquad \dot{\phi}(x,0)
= A \sum_{j=1}^{j_{\rm max}} \cos\left(2 \pi j \frac{x}{L} - \psi_j\right),
\end{equation}
where the $\psi_j$'s are random phases and $2 \pi j_{\rm max}/L$ is of the order
$m$. We will average over several runs, typically 10-20.

For simplicity we will take the quantum modes equal to the zero-temperature
vacuum form: 
\begin{equation}
n_\alpha^0 = 0, \qquad f_k(x,0) = \frac{{\rm e}^{i k x}}{\sqrt{2 \omega_k L}},
\qquad
\dot{f}_k(x,0) = - \frac{i \omega_k {\rm e}^{i k x}}{\sqrt{2 \omega_k L}},
\end{equation}
with $\omega_k = \sqrt{m^2 + k^2}$, $m$ chosen as the zero-temperature
mass and $L$ the size of the system.
To put the system on a computer we discretise it on a space-time lattice. In the
limit of (spatial) lattice spacing $a$ to zero, the mode sum $\Sigma$ occurring
in the equations of motion is logarithmically diverging, but this can be fixed
by a simple mass renormalisation:
$\mu^2 \rightarrow \mu_{\rm ren}^2 \equiv \mu^2 - \frac{3 \lambda}{4 \pi}
\log(\lambda a^2)$.

\paragraph*{Thermalisation}
If $\lambda$ is small and the system is not too far from equilibrium the
interaction of the mode functions with the typically \emph{inhomogeneous} mean
field can be viewed as scattering of particles via the mean field. Because of
this scattering they will hopefully thermalise according to a thermal -- i.e.
Bose-Einstein -- distribution.

To check this, a definition of particle number is needed. For a free
thermal system the 2-point functions read:
\begin{equation}
\langle \hat{\phi}_x \hat{\phi}_y \rangle_{\rm conn} =
\sum_k \frac{1}{\omega_k} \, \frac{n_k + \frac{1}{2}}{L} \, {\rm e}^{i k (x-y)},
\qquad
\langle \hat{\pi}_x \hat{\pi}_y \rangle_{\rm conn} =
\sum_k \omega_k \, \frac{n_k + \frac{1}{2}}{L} \, {\rm e}^{i k (x-y)}.
\label{eq:2pnt}
\end{equation}
We want to define an $n_k$ and $\omega_k$ in a similar way as Eq.~\ref{eq:2pnt}.
This can be done by coarsening the measured 2-point functions over the initial
conditions, a time interval, and space.

\section*{Numerical results}
\begin{figure}
\epsfig{file=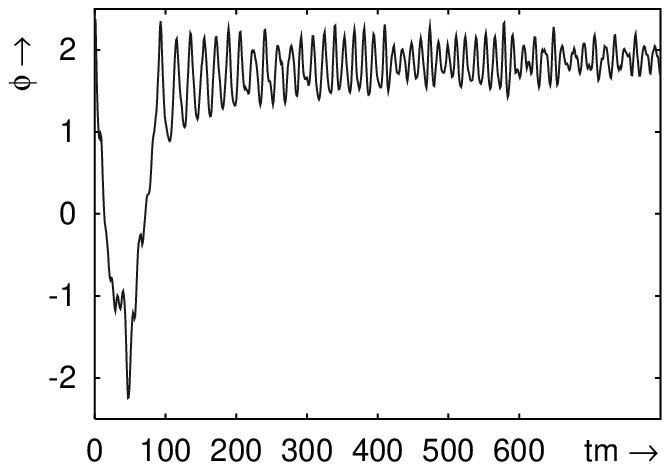}
\epsfig{file=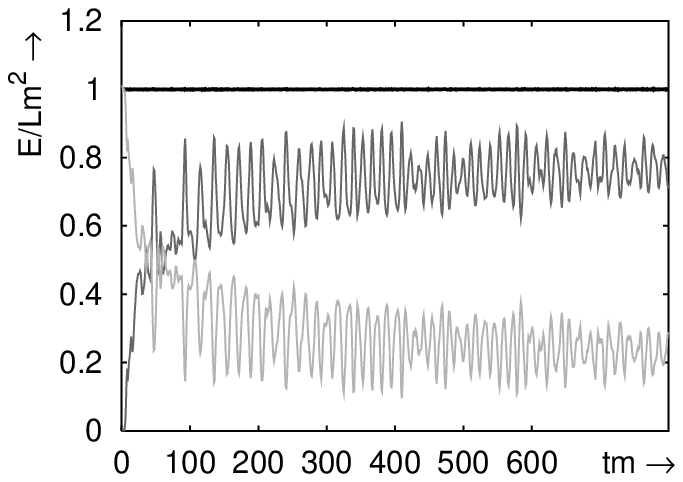}
\caption{Left: time dependence of the average mean field. Right: the energy
densities for the same run: total (straight line at 1), mean field
(starting at total energy), and modes.} \label{fig:phi}
\end{figure}
In Fig.~\ref{fig:phi} we plotted for a typical run ($256$ lattice points, size
$Lm = 32$, $\lambda/m^2=1/12$) the spatial averages of the mean field and the
energy densities as a function of time. We see that the energy, initially
completely in the mean field, is going to the modes with a timescale of the
order 100 in mass units. The total energy is conserved up to a few percent. The
energy staying in the mean field -- around 20\% -- is much more than expected
from classical equipartition: around $1/256$ for $256$ mode functions.
\begin{figure}
\epsfig{file=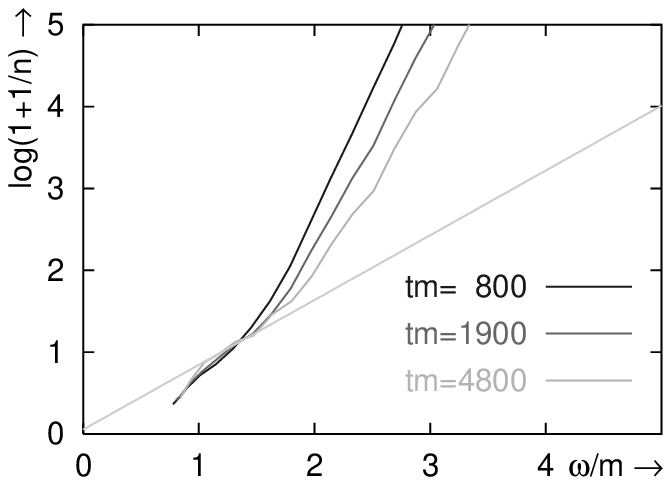}
\epsfig{file=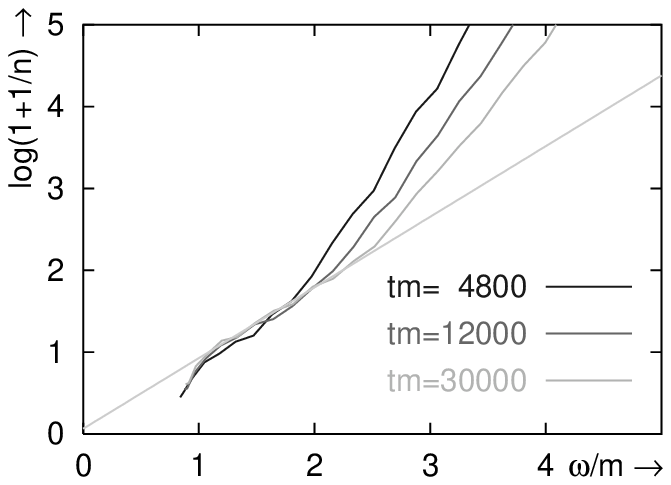}
\caption{The particle numbers for smaller and longer times. For a true BE
$\log(1+1/n)=\beta \omega$. The straight line is a BE fit at the latest time
plotted.} \label{fig:distrib}
\end{figure}
Fig.~\ref{fig:distrib} shows a plot of the calculated particle numbers
obtained after averaging over whole space, approximately 5 oscillation periods
and 10 configurations. On the left the situation is plotted at shorter times, on
the right at larger times. We see that starting at low momenta, a Bose-Einstein
distribution is emerging. Only the zero mode is showing deviations.

Runs at twice this energy density show very similar results, but with faster
thermalisation. At large times deviations from the straight line BE's start
to show up, which we interpret as signs of classical equipartition. For strong
coupling or after a long times we see the emergence of an offset, which might be
a chemical potential following from approximate particle conservation.

We finally mention checks with different kinds of initial conditions (not
presented here), which also show thermalisation according to a Bose-Einstein
distribution. 

\section*{Conclusion}
Using our Hartree ensemble approximation with inhomogeneous mean fields, we find
thermalisation according to a Bose-Einstein distribution, starting in the low
momentum modes. The time-scales heavily depend on the size of the coupling and
the energy density.

Signs of classical equilibration occur only at timescales much larger than that
for the onset of the Bose-Einstein thermalisation, so even if they will prevent
the BE thermalisation to reach arbitrary high values of $\omega$, they do not
influence processes with an intermediate timescale.

An extension to higher dimensions is wanted but numerically demanding. A
possible way out is a reduction of the number of mode functions.

A more elaborate discussion of these simulations is in preparation.

\end{document}